\documentclass[aps,preprint]{revtex4-2}
\usepackage[utf8]{inputenc}
\usepackage{geometry}
\usepackage{amsmath}
\usepackage{amsfonts}
\usepackage{amssymb}
\usepackage{xcolor}
\usepackage{dsfont}
\usepackage{color}
\usepackage{graphicx}
\usepackage{braket}
\usepackage{amsmath}
\usepackage{tikz}
\usepackage{mathdots}
\usepackage{yhmath}
\usepackage{cancel}
\usepackage{color}
\usepackage{siunitx}
\usepackage{array}
\usepackage{multirow}
\usepackage{amssymb}
\usepackage{textcomp}
\usepackage{gensymb}
\usepackage{tabularx}
\usepackage{extarrows}
\usepackage{booktabs}
\usetikzlibrary{fadings}
\usetikzlibrary{patterns}
\usetikzlibrary{shadows.blur}
\usetikzlibrary{shapes}
\usepackage{slashed}
\usepackage{float}
\usepackage{braket}
\usepackage{color}
\usepackage{upgreek}
\usepackage{hyperref}

\hypersetup{
    colorlinks,
    linkcolor={blue!50!red},
    citecolor={orange!70!red},
    urlcolor={blue!80!black}
}

\usepackage{amsfonts}
\makeatletter

\begin{document}

\begin{titlepage}
\begin{flushright}
    Dated: March 31, 2023
\end{flushright}    
    \title{\Large Holographic Quantum Gravity and Horizon Instability}
    \author{Vaibhav Kalvakota \\ \textit{Turito Institute, 500081, Hyderabad, India} \vskip 0.4in {\small{\vskip -1cm
            {\textsf{Email: vaibhavkalvakota@icloud.com}}}}}
\vspace{0.50in}
\begin{abstract}
In this Essay, we will look at the relation between the No Transmission principle and the Strong cosmic censorship (SCC), which we will highlight in the background of quantum gravity. We show that taking quantum gravity into account, one can provide a complete picture of the instability of the inner horizon and the principle that two independent CFTs, under the gauge-gravity duality, imply that the dual bulks must also be independent in that there must not exist a way to transmit a signal between the two spacetimes. We show that this can simply be interpreted as SCC, and that the inner horizon must be unstable (at either linear or nonlinear orders) to be in accordance with holographic quantum gravity.
\end{abstract}
\maketitle
\vskip 0.85in
\begin{flushleft}
\textit{Essay written for the Gravity Research Foundation 2023 Awards for Essays on Gravitation.} \end{flushleft}
\end{titlepage}

Rotating and charged black holes have a characteristic feature of an inner horizon, which has been the core of a debate on the deterministic properties of general relativity, on the basis of the \textit{strong cosmic censorship} conjecture (SCC) (see recent works such as \cite{Dafermos:2017dbw, Dias:2018ynt} ). The attention has been from both the classical as well as the semiclassical and quantum regimes, where such a horizon has some interesting properties. In particular, the nature of the stability of the inner horizon in the semiclassical regime is rather intriguing, since in certain cases, SCC can be violated in the classical limit but preserved in the semiclassical regime (see for instance, \cite{Hollands:2019whz, Dias:2018etb, Dias:2018ynt, Emparan:2020rnp}). This is of interest particularly since one would expect that SCC is preserved in cases where it would \textit{apparently} be violated classically as we will provide an outlook below. Classical violations are well known, and have been studied particularly under two formulations of the SCC: the $C^{2}$ formulation, where curvature divergence at the Cauchy horizon $\mathcal{CH}_{\mathcal{I}}$ implies a non-$C^{2}$ nature of the spacetime, and the \textit{Christodoulou} formulation, where the maximal development for a spacetime with (locally) square-integrable Christoffel symbols cannot be beyond $\mathcal{CH_{I}}$. Another formulation is that of a $C^{0}$ version, which we will defer to discuss here. The $C^{2}$ framework is not a convenient formulation, since the tidal forces experienced by an infalling observer would be finite, allowing it to continue. Versions of SCC in the case of finite energy-fluxes across $\mathcal{CH}_{\mathcal{I}}$ have been found in dS, see for instance \cite{Chambers:1994ap} in Kerr-dS spacetime. Presently, our framework would be that of the semiclassical Einstein field equations, given by 
\begin{equation}
	G_{\mu \nu }[g, \Lambda ]=\langle T_{\mu \nu }\rangle .
\end{equation}
We will discuss how quantum gravity allow one to preserve SCC even if it is violated classically, for instance in the case of the BTZ black hole, and how this directly implies that the holographic quantum gravity description of the independence of two CFTs can be found \cite{Engelhardt:2015gla}. However, we will first motivate the point of this essay: \textit{should SCC be true? What are the implications if we let it be violated?} Let $V$ be a null component in the Kruskal coordinate system. Then, the renormalized stress-tensor must diverge at the inner Cauchy horizon $\mathcal{CH}_{\mathcal{I}}$ so as to prevent SCC from being violated. This is understood as follows: taking into consideration the semiclassical description of gravity, the geometry of $(M, g)$ is described by some metric $g$, with the matter fields description being put forward by the stress-tensor $\langle T_{\mu \nu }\rangle $. However, due to backreactions, one must take into account of the perturbative analysis of the metric taking backreactions upto successive orders to find the description of the quantum backreactions and their effect on the geometry of the spacetime. A way of preserving SCC in semiclassical gravity has been put forward in \cite{Hollands:2019whz} and considered in papers such as \cite{Dias:2019ery, Emparan:2020rnp}, which shows that the behaviour of $\langle T_{\mu \nu }\rangle $ affects the inner horizon by behaving singularly, therefore preserving SCC. For a free scalar field in the null Kruskal components, we would have 
\begin{equation}\label{eq:sgst}
	\langle T_{VV} \rangle  \sim \frac{C}{V^{2}},
\end{equation}
where $C$ is a constant based on the parameters describing the black hole. For a charged AdS black hole, this would be in terms of $M$ and $q$ and the coordinate $V\to 0^{-}$ at the inner horizon. The hope is that due to the singular nature of the stress-tensor $\langle T_{VV} \rangle $, $\mathcal{CH}_{\mathcal{I}}$ becomes unstable, preserving SCC.

This is important in a particularly interesting aspect: let us assume a signal packet could be sent across the inner horizon, ignoring one of the core issues, where the complete history of the outer cosmology could be viewed in a definite proper-time. Then, this packet could emerge in an external spacetime, and implies causal transmission between two otherwise disconnected spacetimes. As per the \textit{No Transmission principle} \cite{Engelhardt:2015gla, Engelhardt:2016kqb} one should not be able to send this wavepacket from one AdS spacetime to another AdS spacetime (or indeed any spacetimes following holographic quantum gravity) which are otherwise disconnected, and in the boundary perspective one should not be able to send this packet between the CFTs dual to these two bulk spacetimes. This is akin to the prediction of SCC; one should not be able to traverse through $\mathcal{CH}_{\mathcal{I}}$ in the first place, to ensure that two disconnected spacetimes do not have a causal connectedness, and all the same to address the blue-shift problem. Could one make some statement in a quantum gravity theory about $\mathcal{CH}_{\mathcal{I}}$? For instance, consider a charged AdS black hole: one could, in principle, send a wavepacket originating from some $CFT_{B}$ for the initial AdS spacetime $AdS_{B}$, and let the wavepacket cross the inner horizon. This could theoretically reach some $CFT_{A}$ of some $AdS_{A}$, and violate the No Transmission principle. Therefore, the stability of $\mathcal{CH}_{\mathcal{I}}$ plays a crucial role in determining whether or not the No Transmission principle holds. 

\begin{figure}
	\centering

\tikzset{every picture/.style={line width=0.75pt}} 

\begin{tikzpicture}[x=0.75pt,y=0.75pt,yscale=-0.90,xscale=0.90]

\draw   (116.79,60.18) -- (212.22,166.76) -- (116.79,273.34) -- (21.37,166.76) -- cycle ;
\draw   (301.4,20.03) -- (390.61,119.91) -- (301.4,219.79) -- (212.2,119.91) -- cycle ;
\draw    (180.23,130.42) -- (180.3,202.5) ;
\draw    (244.02,84.37) -- (244.27,156.29) ;
\draw    (212.22,123.14) .. controls (213.89,124.81) and (213.89,126.47) .. (212.22,128.14) .. controls (210.55,129.81) and (210.55,131.47) .. (212.22,133.14) .. controls (213.89,134.81) and (213.89,136.47) .. (212.22,138.14) .. controls (210.55,139.81) and (210.55,141.47) .. (212.22,143.14) .. controls (213.89,144.81) and (213.89,146.47) .. (212.22,148.14) .. controls (210.55,149.81) and (210.55,151.47) .. (212.22,153.14) .. controls (213.89,154.81) and (213.89,156.47) .. (212.22,158.14) .. controls (210.55,159.81) and (210.55,161.47) .. (212.22,163.14) -- (212.23,166.76) -- (212.23,166.76) ;
\draw    (470.05,6.82) -- (469.91,101.46) ;
\draw    (469.91,101.46) .. controls (471.57,103.13) and (471.57,104.79) .. (469.9,106.46) .. controls (468.23,108.13) and (468.23,109.79) .. (469.89,111.46) .. controls (471.55,113.13) and (471.55,114.79) .. (469.88,116.46) .. controls (468.21,118.13) and (468.21,119.79) .. (469.88,121.46) .. controls (471.54,123.13) and (471.54,124.79) .. (469.87,126.46) .. controls (468.2,128.13) and (468.2,129.79) .. (469.86,131.46) .. controls (471.52,133.13) and (471.52,134.79) .. (469.85,136.46) .. controls (468.18,138.13) and (468.18,139.79) .. (469.85,141.46) .. controls (471.51,143.13) and (471.51,144.79) .. (469.84,146.46) .. controls (468.17,148.13) and (468.17,149.79) .. (469.83,151.46) .. controls (471.49,153.13) and (471.49,154.79) .. (469.82,156.46) .. controls (468.15,158.13) and (468.15,159.79) .. (469.82,161.46) .. controls (471.48,163.13) and (471.48,164.79) .. (469.81,166.46) .. controls (468.14,168.13) and (468.14,169.79) .. (469.8,171.46) .. controls (471.47,173.13) and (471.47,174.79) .. (469.8,176.46) .. controls (468.13,178.13) and (468.13,179.79) .. (469.79,181.46) .. controls (471.45,183.13) and (471.45,184.79) .. (469.78,186.46) .. controls (468.11,188.13) and (468.11,189.79) .. (469.77,191.46) -- (469.77,196.1) -- (469.77,196.1) ;
\draw    (469.77,196.1) -- (469.63,290.73) ;
\draw    (593.97,8.25) -- (593.83,102.89) ;
\draw    (593.83,102.89) .. controls (595.5,104.56) and (595.5,106.22) .. (593.83,107.89) .. controls (592.16,109.56) and (592.16,111.22) .. (593.82,112.89) .. controls (595.48,114.56) and (595.48,116.22) .. (593.81,117.89) .. controls (592.14,119.56) and (592.14,121.22) .. (593.8,122.89) .. controls (595.47,124.56) and (595.47,126.22) .. (593.8,127.89) .. controls (592.13,129.56) and (592.13,131.22) .. (593.79,132.89) .. controls (595.45,134.56) and (595.45,136.22) .. (593.78,137.89) .. controls (592.11,139.56) and (592.11,141.22) .. (593.77,142.89) .. controls (595.44,144.56) and (595.44,146.22) .. (593.77,147.89) .. controls (592.1,149.56) and (592.1,151.22) .. (593.76,152.89) .. controls (595.42,154.56) and (595.42,156.22) .. (593.75,157.89) .. controls (592.08,159.56) and (592.08,161.22) .. (593.74,162.89) .. controls (595.41,164.56) and (595.41,166.22) .. (593.74,167.89) .. controls (592.07,169.56) and (592.07,171.22) .. (593.73,172.89) .. controls (595.39,174.56) and (595.39,176.22) .. (593.72,177.89) .. controls (592.05,179.56) and (592.05,181.22) .. (593.72,182.89) .. controls (595.38,184.56) and (595.38,186.22) .. (593.71,187.89) .. controls (592.04,189.56) and (592.04,191.22) .. (593.7,192.89) -- (593.69,197.52) -- (593.69,197.52) ;
\draw    (593.69,197.52) -- (593.55,292.16) ;
\draw  [dash pattern={on 4.5pt off 4.5pt}]  (470.05,6.82) -- (593.83,102.89) ;
\draw  [dash pattern={on 4.5pt off 4.5pt}]  (593.97,8.25) -- (469.91,101.46) ;
\draw  [dash pattern={on 4.5pt off 4.5pt}]  (469.91,101.46) -- (593.69,197.52) ;
\draw  [dash pattern={on 4.5pt off 4.5pt}]  (469.77,196.1) -- (593.83,102.89) ;
\draw  [dash pattern={on 4.5pt off 4.5pt}]  (469.77,196.1) -- (593.55,292.16) ;
\draw  [dash pattern={on 4.5pt off 4.5pt}]  (469.63,290.73) -- (593.69,197.52) ;

\draw (145.26,153.99) node [anchor=north west][inner sep=0.75pt]    {$\mathcal{CH}_{\mathcal{I}}$};
\draw (246.76,108.1) node [anchor=north west][inner sep=0.75pt]    {$\mathcal{CH} '_{\mathcal{I}}$};
\draw (95.69,96.6) node [anchor=north west][inner sep=0.75pt]    {$CFT_{L}$};
\draw (95.91,163.68) node [anchor=north west][inner sep=0.75pt]    {$AdS_{L}$};
\draw (297.68,117.04) node [anchor=north west][inner sep=0.75pt]    {$AdS_{R}$};
\draw (281.63,49.38) node [anchor=north west][inner sep=0.75pt]    {$CFT_{R}$};
\draw (540.8,217.2) node [anchor=north west][inner sep=0.75pt]  [font=\scriptsize,rotate=-322.95]  {$\mathcal{CH}_{O}$};
\draw (501.55,201.96) node [anchor=north west][inner sep=0.75pt]  [font=\scriptsize,rotate=-39.78]  {$\mathcal{CH}_{O}$};
\draw (496.51,181.53) node [anchor=north west][inner sep=0.75pt]  [font=\scriptsize,rotate=-323.03]  {$\mathcal{CH}_{\mathcal{I}}$};
\draw (544.1,167.09) node [anchor=north west][inner sep=0.75pt]  [font=\scriptsize,rotate=-37.57]  {$\mathcal{CH}_{\mathcal{I}}$};
\draw (483,20.4) node [anchor=north west][inner sep=0.75pt]    {$CFT_{B}$};
\draw (545,259.4) node [anchor=north west][inner sep=0.75pt]    {$CFT_{A}$};

\end{tikzpicture}
\caption{(Left) An ‘un-conformal’ schematic of two AdS spacetimes with parameter space $(M, q, L, \Lambda )$. The CFTs dual to the independent bulks cannot be composed in a single joint space. The wiggly line depicts a wavepacket transmitting to $AdS_{B}$ by violating SCC and the No Transmission principle. Therefore, $\mathcal{CH}_{\mathcal{I}}$ must be in accordance with SCC and be unstable. (Right) A conformal diagram of the charged AdS spacetime. $CFT_{A}$ and $CFT_{B}$ are independent and must obey the SCC to also ensure the No Transmission principle by forbidding traversing from the bottom CFT to the top CFT.}
\end{figure}
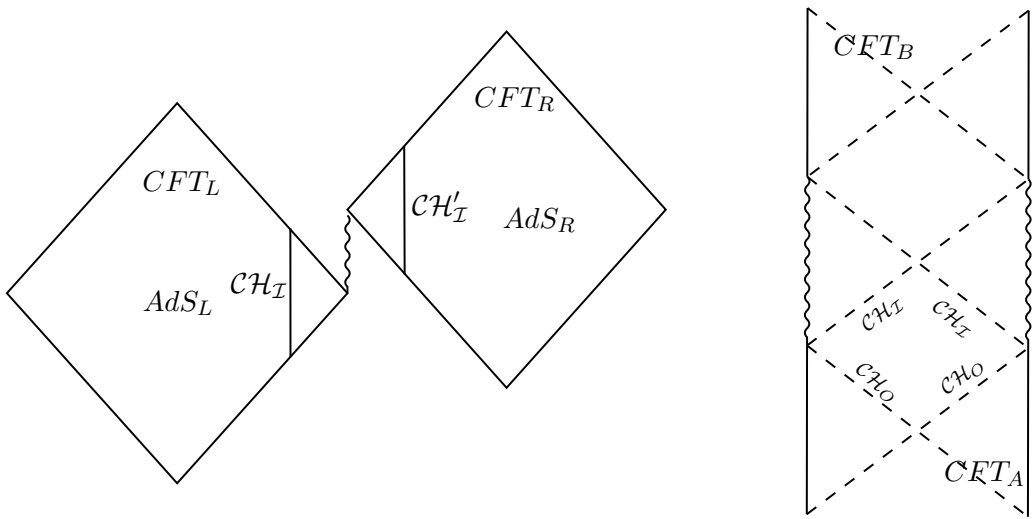

If $\mathcal{CH}_{\mathcal{I}}$ is \textit{indeed} unstable, then it resolves the problem immediately. However, does this also hold in the case of linear quantum backreactions adding to the misery of the problem of stability? One would expect the No Transmission principle to also hold in such cases, but the stability of $\mathcal{CH}_{\mathcal{I}}$ must be taken into account at linear order backreaction. Even for finding stability without backreactions, the way to do this would be using the stress-tensor based condition \eqref{eq:sgst} and look at the behaviour of the stress-tensor. For instance, consider the case of a BTZ black hole in AdS with a rotational parameter. Then, the constant $C\sim M, L$, and one could investigate the behaviour of $\mathcal{CH}_{\mathcal{I}}$, which was found to be divergent in \cite{Emparan:2020rnp} considering nonlinear backreactions but classically violated as shown in the case of rotational BTZ spacetimes; see\cite{Dias:2019ery, Balasubramanian:2019qwk}. 

Let us stick with the case of a BTZ spacetime. In this case, one would expect a violation of SCC \cite{Dias:2019ery}, but semiclassically one could save it using the approach mentioned above, which was carried out in \cite{Emparan:2020rnp} following \cite{Hollands:2019whz}. Naturally, the behaviour of the stability of $\mathcal{CH}_{\mathcal{I}}$ is given by the regularity condition,
\begin{equation}\label{eq:reg}
	\beta =\frac{\Im \omega }{\kappa  },
\end{equation}
where $\kappa $ is the surface gravity at $\mathcal{CH}_{\mathcal{I}}$ and $\alpha \equiv \Im \omega $ is the minimum over the $\Im $ quasi-normal frequencies and is called the spectral gap. Naturally, this regularity condition can be set by identifying specific values of $\beta $. For instance, the work by Moss et al \cite{Mellor:1989ac} showed that considering a scalar field $\Phi $, for values of the parameter $\beta <1$, the scalar field behaves particularly badly by not being $C^{1}$, due to which the derivatives also behave badly at $\mathcal{CH}_{\mathcal{I}}$, preserving the $C^{2}$ formulation of SCC, although for other $\beta $ values with near $|Q|\sim M$ Reissner-Nordstrom-dS black holes do not obey this condition. 

A continuation for the formulation of \eqref{eq:sgst} is to add in another factor $t_{VV}$ \cite{Hollands:2019whz}:
\begin{equation}
	\langle T_{\mu \nu }\rangle \sim \frac{C}{V^{2}}+t_{VV}.
\end{equation}
In the case where $C=0$, one would expect $t_{VV}$ to diverge, and when $C\neq 0$ $\langle T_{\mu \nu }\rangle \sim \frac{C}{V^{2}}$ should diverge. This would imply large backreactions, to the geometry of $\mathcal{CH}_{\mathcal{I}}$, which would preserve SCC under specific $\beta $ settings on $t_{VV}$. In fact, one could use this formulation to precisely show in the Reissner-Nordstrom-dS spacetime that since the behaviour of $\langle T_{\mu \nu } \rangle $ is very bad, SCC would be saved. In the $D=3$ case of BTZ spacetime where $C=0$, the backreactions could be taken at nonlinear orders, which show that SCC does hold, even if not at a linear level \cite{Emparan:2020rnp}. In general, SCC can be interpreted as a formulation of the No Transmission principle; if SCC does not hold, it must also be that the wavepacket could instead emerge in an external asymptotic region. 

Even if in terms of plain QFTs, one would expect that the No Transmission principle holds, instead of just the particular case of AdS/CFT, which we used to highlight a particular example. A particularly nice way of imagining this is to look at these (conformal) QFTs as CFTs mapped into different spacetimes; if there would be a causal connection naturally, then they could be mapped into one spacetime, but if the No Transmission principle holds and these CFTs are independent, they must be mapped into two different causally disconnected spacetimes. In terms of a Hilbert space decomposition, this independence would be illustrated with the tensor product
\begin{equation}
	\mathcal{H}=\mathcal{H}_{A}\otimes \mathcal{H}_{B}.
\end{equation}
Gauge-gravity duality suggests that since the CFTs are independent, the bulk duals must also be independent, which is what we used to provide an outlook for the strong relation between SCC and the No Transmission principle -- since the two bulks must also be independent, one could illustrate the preservation of the No Transmission principle via SCC, which destabilises the inner horizon and prevents any wavepacket from emerging in some independent bulk. Classically, there may be violations of SCC, but one could invoke the results discussed above to say that quantum backreactions could preserve SCC and guarantee the No Transmission principle in spacetimes where SCC is invoked. 

It is therefore interesting, to see where quantum gravity suggests some support to problems that might be classically problematic. In the case of SCC, we have two results: (1) quantum backreactions could save SCC at higher orders, and (2) the SCC in turn saves the No Transmission principle. As an example, the quantum backreactions that allow the Reissner-Nordstrom-dS spacetime to stay in accordance with SCC would automatically imply the No Transmission principle. Due to this, one could infer that holographic quantum gravity does not allow one to resolve the singularity in a way that the No Transmission principle is violated, and provides backreactions strong enough to ensure the stress-tensor diverges, so as to preserve SCC. This is summed up in the following:

\textbf{Conjecture}: \textit{In the classical regime, SCC could be violated. But one could conjecture that quantum gravity preserves SCC (at linear or nonlinear orders) by causing a divergence in the stress-tensor, which would backreact to the geometry to cause curvature blow-ups. This would imply that quantum gravity would correct classical violation of SCC to preserve the holographic No Transmission principle}.

One could be a little more adventurous and try to find a relation between quantum gravity and the stability of the outer horizon, which naturally generalizes to any uncharged and unrotating black hole spacetime as well. However, this would not be of much use, since it is well known that in $D\geq 5$ the Gregory-Laflamme instability poses a violation of the weak censorship conjecture, by developing singularities on the ‘pinched’ horizon. This could be a \textit{weak} violation of WCC, however this result would not be as strong as SCC for the same reason as in classical gravity; the exterior of the black hole may or may not be safe, i.e. $\mathcal{I}^{+}$ could be causally disconnected from the black hole region $\mathcal{B}=M\backslash J^{-}(\mathcal{I})$, but the deterministic nature of GR is a different issue than this. The validity of SCC is therefore mathematically independent of WCC. Due to this, a quantum gravity description of WCC might be of interest on its own, but with regard to the No Transmission principle with regard to gauge-gravity duality, it might not be of much significance.

\textbf{Acknowledgements:} I thank Aayush Verma for helpful comments and discussions. I am grateful to ICTS-TIFR for providing stay during the initial parts of this work. 

\bibliography{ref.bib}
\end{document}